\documentclass[fleqn,twoside]{article}
\usepackage{espcrc2}

\usepackage{graphicx}
\usepackage[figuresright]{rotating}

\def\be{\begin{eqnarray}&&}

 \def\ee{ \end{eqnarray}}

\newcommand{\AmS}{{\protect\the\textfont2
  A\kern-.1667em\lower.5ex\hbox{M}\kern-.125emS}}

\title{Theoretical Descriptions of the Nucleon 
Electromagnetic Form Factors 
in the Space- and
Time-like regions}

\author{Giovanni Salm\`e\address{Istituto Nazionale di Fisica Nucleare
- Sezione di Roma \\ 
        P.le A. Moro 2, 00185 Rome, Italy}}
       
\begin{document}

\begin{abstract}
Some recent advances  in the theoretical description of the 
Nucleon
electromagnetic form factors,  both  in the space- and time-like regions, will be 
briefly illustrated.  In particular,  both
the present stage of the lattice calculations  and  updated phenomenological
approaches, like the ones based on dispersion relations and on microscopical
models, will be reported.

\vspace{1pc}
\end{abstract}

\maketitle

\section{Introduction}
The Nucleon electromagnetic form factors (em ff's) represent 
a relevant chapter in our path
towards the understanding of hadron dynamics. It is striking that,  after
almost  half a century of  experimental investigations, this field  
could still hold a number of  surprises (e.g., the intriguing fall-off of the proton charge 
ff more rapid than
the assumed dipole one), or    have some kinematical
regions not fully explored (e.g.,  
 the timelike region, in order to obtain  separate information on 
ff's ). 
  In particular, let us remind (see, e.g., \cite{arrington,dejager} 
for recent reviews) that in the spacelike (SL) region 
 the existence of a possible zero in  $G_E^p$ for $-q^2=Q^2> 7~GeV^2$ ($q^2=$ the square momentum transfer), 
 as suggested by a linear extrapolation of the data obtained by the polarization
 transfer experiments,  could open  new 
perspectives. Indeed, the
dipole form of the  Nucleon ff's is supported by the 
pQCD, therefore 
a sharp difference between the charge distribution of the current quarks and 
the
charge distribution of the "constituent" quarks appears a clear signature of
NpQCD effects in act. 
 Even low-$Q^2$ ff's (cf the forthcoming analysis of E08-007 at TJLAB) 
 could give new stringent constraints for
builders of phenomenological models.
In the the timelike (TL) region, analogously relevant issues can be met. For
instance,
the neutron data around $q^2 \sim 4 ~GeV^2$  show a  behavior, 
$|G^n_{eff}|\sim |G^p_{eff}|$, sharply different from the
one suggested by pQCD, naively indicating 
$|G^n_{eff}|\sim |e_d/e_u|~|G^p_{eff}|$. Moreover,
 TL proton data show many interesting structures (see, e.g. \cite{babar}), 
 that could give a lot of
 information on the hadronic component of the photon wave function.  
 
 On the theoretical side, in a very schematic way, one can list the following 
 approaches:
 i) Lattice calculations, based on a Euclidean 4D space,  that can address
   the  SL region
  only; ii) 
analytic approaches, based on dispersion relations, that can investigate both
 SL and TL  regions, even yielding  predictions for the unphysical interval 
 $0<q^2<4M^2_N$; iii) purely phenomenological methods, with 
 constituent quarks (CQ) as basic ingredients, that can be exploited for
  describing both SL and
 TL regions; iv) field theoretical approaches, like the ones  based on   the  
Dyson-Schwinger equation, that can be applied to   SL and  TL regions. In this short review we will restrict the
presentation  to  i), ii) and  iii) (see, e.g.,\cite{arrington}
 for a different discussion from i) to iv)). 
 It should be also mentioned a quite
 recent approach based on the gravity dual
models of QCD (or 
holographic QCD), like the one in Ref. \cite{hQCD}.

\section{ Lattice "Data" }
In principle,  Lattice QCD allows one to achieve a non perturbative 
description of
the Nucleon  ff's, as shown by the more and more accurate results
 obtained in  recent years.
In particular, such interesting outcomes have taken profit from: 
i) an impressive growth  of the available computing power, that has allowed 
in nowadays results,
e.g., to
release the {\em quenched approximation} (where  the sea-quark loop
contributions are
disregarded); ii) new algorithms, leading to a 
better description of the 
fundamental symmetries
of QCD,   and  last, but not 
 least,
 iii) substantial improvements of the chiral extrapolation techniques.
 
 On the other hand, we should mention the main obstacles to be overcome:
 \begin{itemize} 
 
 \item 
 in a 4D Euclidean Lattice (i.e., $it\to t$), one cannot  directly evaluate
   the matrix
elements of the em current operator. Therefore 
a careful study of the proper ratio between three- and two-point functions is
necessary in order  to get  "large values" of the Euclidean time, that
eventually allows to
 filter the 
contribution from hadronic states with  Nucleon quantum numbers. 
This
filtering prevents the study of  TL ff's, since an  exponential fall-off 
in time lacks for $q^2>0$;
\item present Lattice data are obtained with values of the coupling constant
$\beta=6/g^2$ and the hopping parameter, $\kappa$, that lead to
   sizable $u,d$ quark
 masses. Usually, the calculations are labeled by the corresponding pion mass,
 from which one can infer  the quark mass in the actual  data. 
 In order to move   toward the current quark masses, one has to
  reduce the Lattice
 spacing (that allows to accommodate lighter quarks) or  to exploit  
 extrapolation formulas, obtained from 
  the 
 low-energy regime of
 the QCD. Unfortunately such   chiral extrapolations have  a well-known 
  "non analytic" ($log~ m^2_\pi$) behavior for small value of the pion mass.
The lowest value of the pion mass in the present Lattice calculations for
the Nucleon ff's is $\sim ~360~MeV$ \cite{LHPC};
\item the  smallness of  momentum transfers  investigated  in a   Lattice is
related to
 the space 
extent, $L$, namely $|\vec q|\sim 2\pi/L$. This has influence on the range of
applicability of the chiral extrapolations, about
$Q^2\le 0.4 ~GeV^2$,  not always reached by the present Lattices  (e.g., 
$Q^2_{low}\sim
0.23 ~GeV^2$ for the LHP 
Coll. \cite{LHPC} and , $Q^2_{low}\sim
0.4 ~GeV^2$ for the QCDSF Coll. \cite{QCDSF} and  $Q^2_{low}\sim
0.15 ~GeV^2$, for the Cyprus-MIT Coll. \cite{Cyprus}, but with  quenching
approximation);
\item  "large" momentum transfers in the range of few $GeV^2$ are in principle accessible by nowadays Lattice spacing, but Fourier transforms of two- and three-point functions become
noise-dominated for square momentum transfers beyond $2~GeV^2$; 
   \end{itemize}

\begin{figure}[t]
\includegraphics[width=6.7cm] {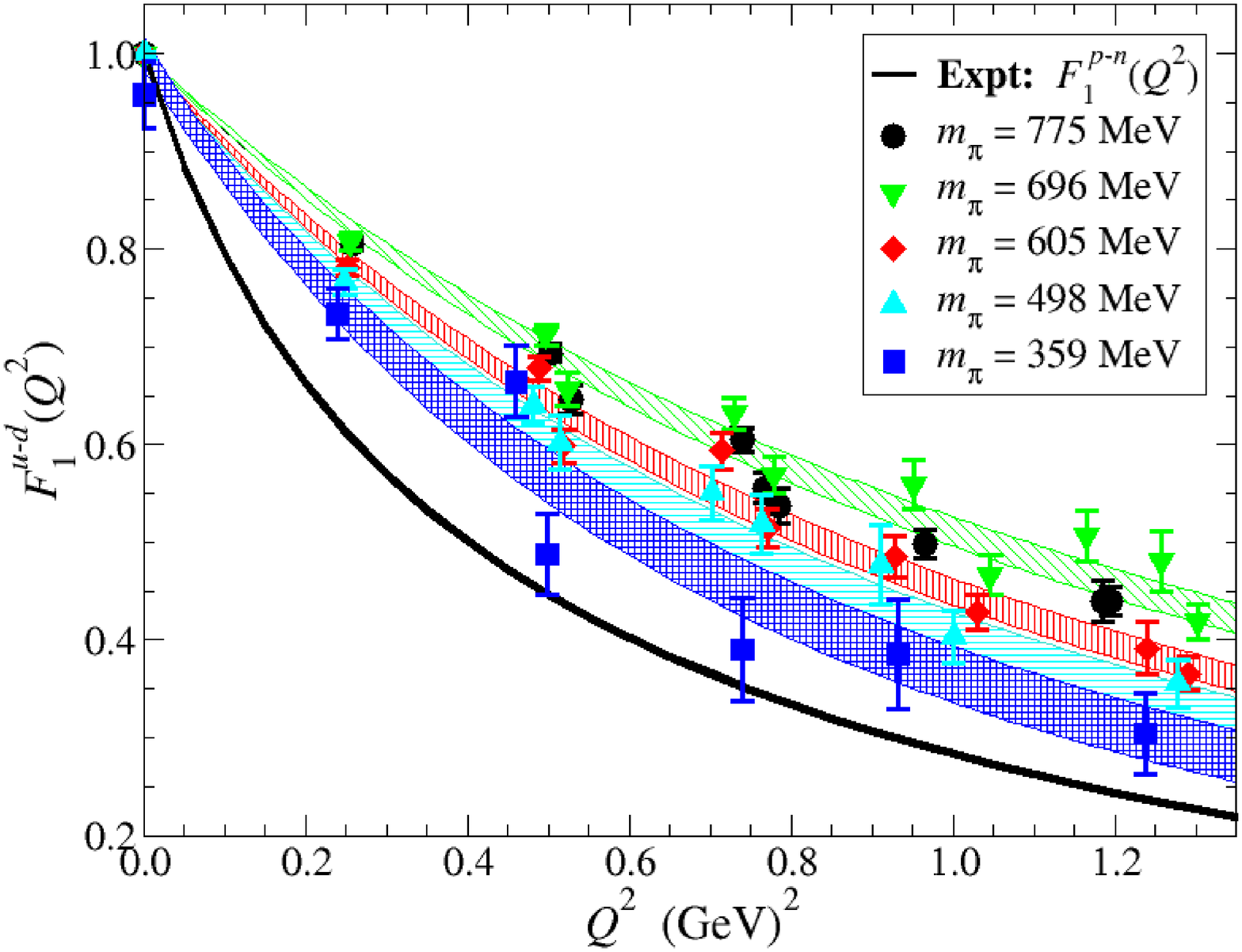}
\includegraphics[width=6.7cm] {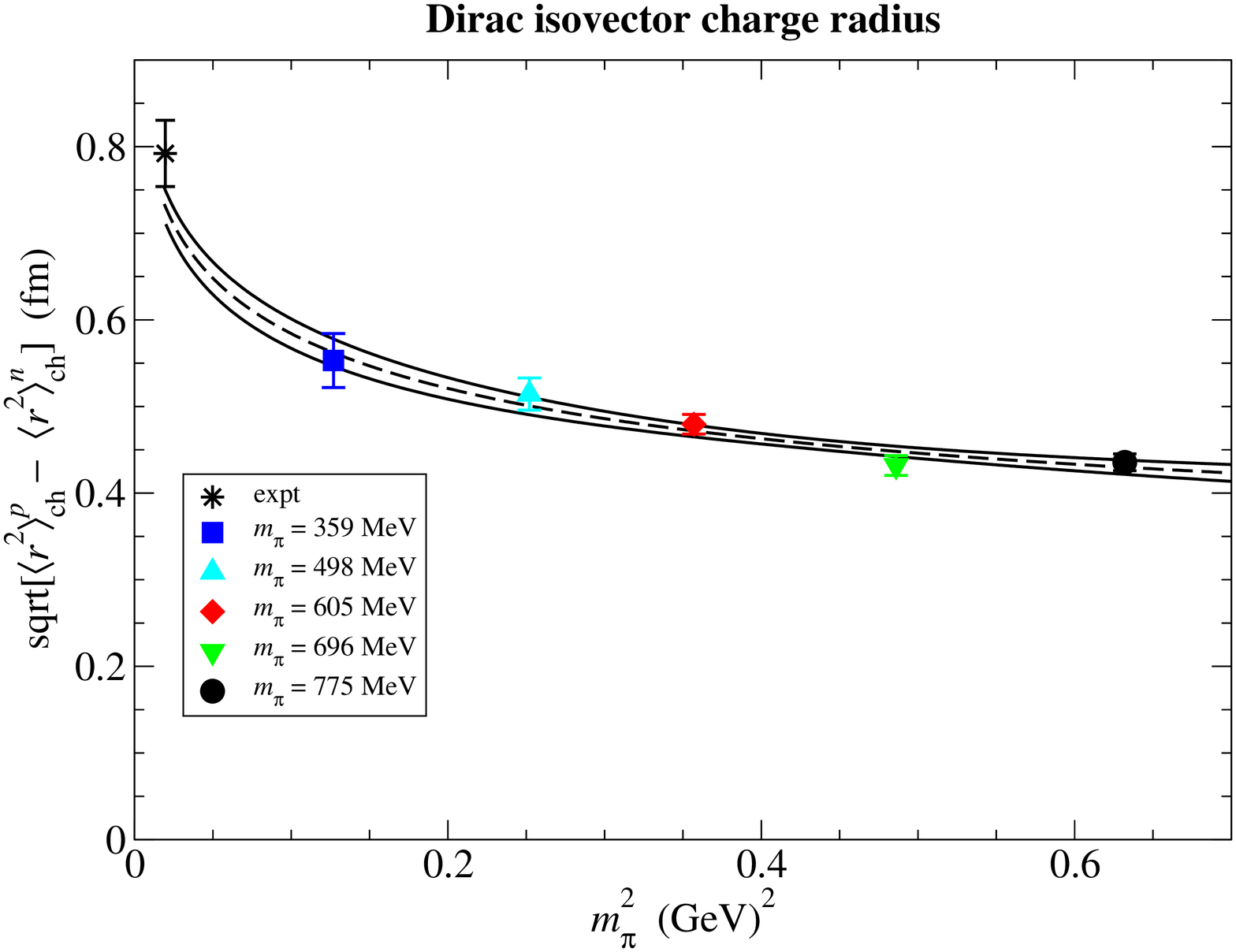}

\caption{ LHP Coll.
results for the  "isovector" (see text) approximation of   the
Dirac ff $F_1(Q^2)$ (upper panel) and  for the 
"isovector" radius (lower panel). Solid and dashed lines: different
extrapolation formulas. (After Ref. \cite{LHPC}).} \label{lhpc}\end{figure}
As an example of an updated Lattice calculation of the Nucleon ff's, 
the results obtained by  the LHP Coll. 
\cite{LHPC}, for both  the "isovector" Dirac ff (obtained through
  $\langle proton| \bar u \gamma^\mu u
- \bar d \gamma^\mu d |proton \rangle$)  and  the "isovector" radius (with
different extrapolations), are shown in Fig. \ref{lhpc}.
\section{ Phenomenology I:  Dispersion Relations }

\begin{figure}[t]
\includegraphics[width=3.5cm, angle=-90] {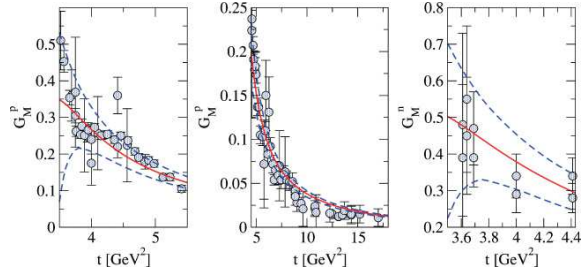}

\caption{ Nucleon ff's for TL momentum
transfers. (After  Ref. \cite{meissner}) } \label{bonn}
 \end{figure}

 Dispersion relation (DR) approaches have a very long story. Within such a
framework, one can relate 
real and imaginary parts of the 
 em ff's, once  the unitarity (probability conservation) and 
 analyticity
 (causality) properties are implemented. 
 The analysis based on DR allows one to investigate both SL and TL regions, 
 and even
 the unphysical region in the TL region, where  a
 possible bound state $p\bar p$ is present, close to  the threshold,
 $q^2=4M_N^2$.
 The main ingredient is the DR given by ($i=1,2$)
 $$ F_i(q^2) = {1\over \pi} \int _{q^2_0}^\infty {Im F_{i}(q^{\prime 2})\over 
 q^{\prime 2} -q^2 -i
 \varepsilon} ~dq^{\prime 2}$$ or a once subtracted version, where $F_{i}(0)$ (if
 finite)
 appears
 $$ F_{i}(q^2) = F_{i}(0)+{q^2\over \pi} \int _{q^2_0}^\infty 
 {Im F_{i}(q^{\prime 2})\over q^{\prime 2}
 \left(q^{\prime 2} -q^2 -i
 \varepsilon \right)} ~dq^{\prime 2}$$
 The next ingredient is a suitable Ansatz,  physically motivated, 
 for $Im F(t^\prime)$. For instance, the Bonn group \cite{meissner} has   exploited 
 i) the form suggested by the decay processes $\rho \to \pi \pi$,
 $\phi \to K \bar K$, $\omega \to \rho \pi$, and related continua and ii) the 
 constrains from pQCD.
In this case, the fit to the
experimental data (SL nucleon 
 ff's and TL proton data), has 17 free parameters and a total 
 $\chi^2/DOF$ of
1.8. It should be pointed out that the TL neutron calculation represents a
prediction of the model. In Fig. \ref{bonn}, the TL Nucleon ff's obtained 
 in the so-called super convergent approach (imposed by a proper fall-off of $F_i$) are shown.

\begin{figure}
\includegraphics[width=3.9cm, angle=-90] {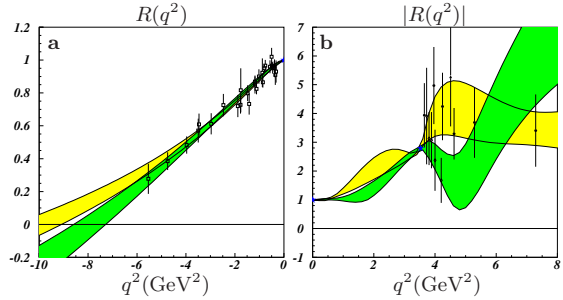}

\caption{ $R(q^2)$ (see text) vs $q^2$. The bands represent the error.
 Green band: retaining in the fitting procedure 
 the TL LEAR data \cite{lear}.
   Yellow band: the same as the green band, but retaining the 
   TL BABAR data \cite{babar}. (After Ref. \cite{pacetti}).
} \label{frascati}\end{figure}

 In Ref. \cite{pacetti} a different approach has been followed, by  directly 
 applying DR  to  the ratio $R(q^2)=\mu_p
G^p_E(q^2)/G^p_M(q^2)$,  both in SL and TL regions. A very  general form for 
$Im~ R(q^2)$, with six parameters,  has been adopted, obtaining 
 a $\chi^2/DOF \sim =1.3$. In Fig. \ref{frascati}, the SL and TL ratios  are
 shown (for  an analysis 
of all the 
ff's
   see  Ref. \cite{baldini}).
The approach   yields  two interesting predictions: i) the position 
of the SL zero, that depends
upon the TL data chosen  in the fit, i.e.
$q^2_z(BABAR) =-10\pm 1~GeV^2$, $q^2_z(LEAR) =-7.9\pm 0.7~GeV^2$; ii) 
the asymptotic limits, extracted from the
theoretical expressions 
$$
\lim_{q^2\to \pm \infty} \left | {G^p_E(q^2) \over G^p_M(q^2)}\right |=
 \left \{
\begin{array}{c}{ 0.95 \pm 0.20 ~ BABAR} \\ { 2.3 \pm 0.7 ~ LEAR}\end{array} 
\right.
$$

\section{ Phenomenology II:
  Microscopic Approaches }

 Relativistic CQ models  have been widely adopted for obtaining  
 a microscopic description of the
 em properties of the Nucleon.  
  Since '90, it was recognized the relevance of dressing the CQ's,
   e.g. by suitable
 ff's, with a spatial extension of the order of 0.4 - 0.5  fm, 
 (a scale
 different from a naive pion-cloud model), in order to obtain a very accurate
 description of the Nucleon and Pion ff's in the SL region (see, e.g.,
 \cite{rome}). After that, new approaches, aimed at a more deep understanding of
 the degrees of freedom beyond the CQ ones, have been developed. 
  
\begin{figure}[t]
\parbox{7.5cm}{\includegraphics[width=3.5cm,angle=-90] {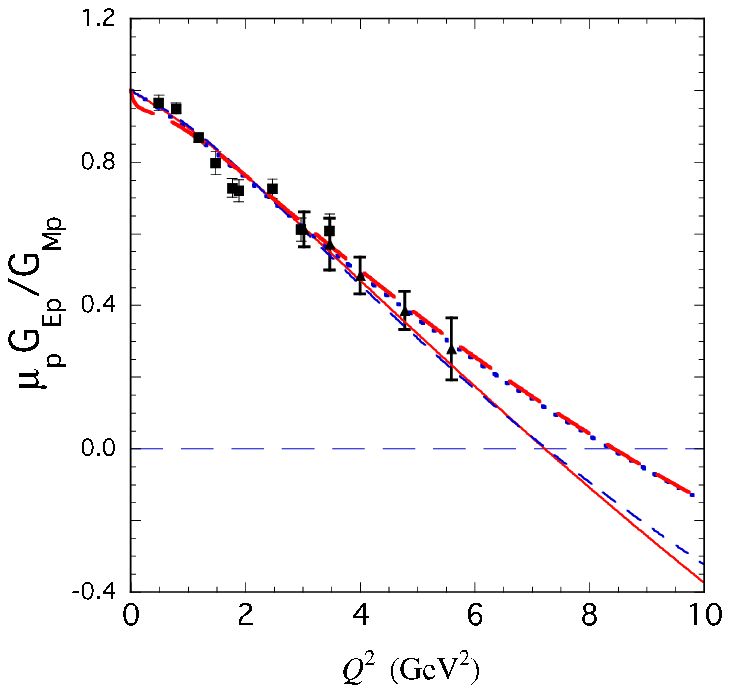}
\includegraphics[width=3.5cm,angle=-90] {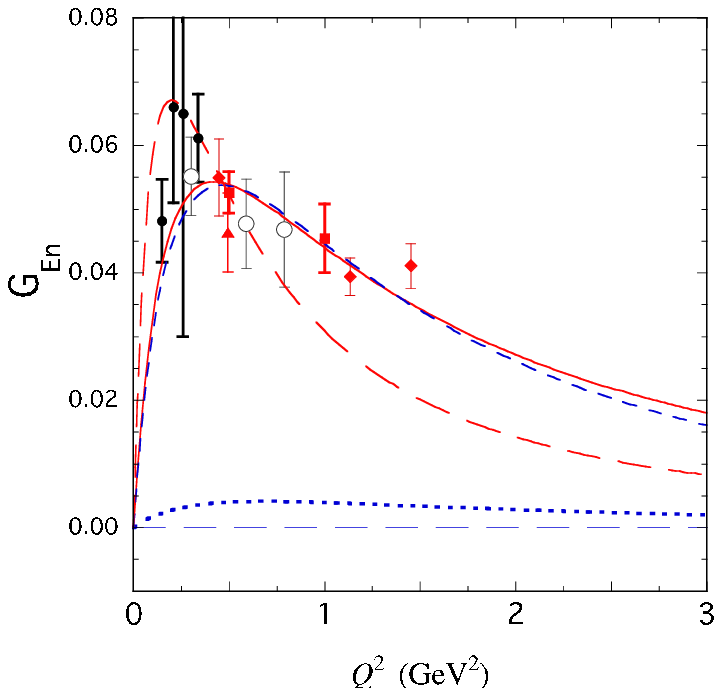}}
\parbox{7.5cm}{\includegraphics[width=3.5cm,angle=-90] {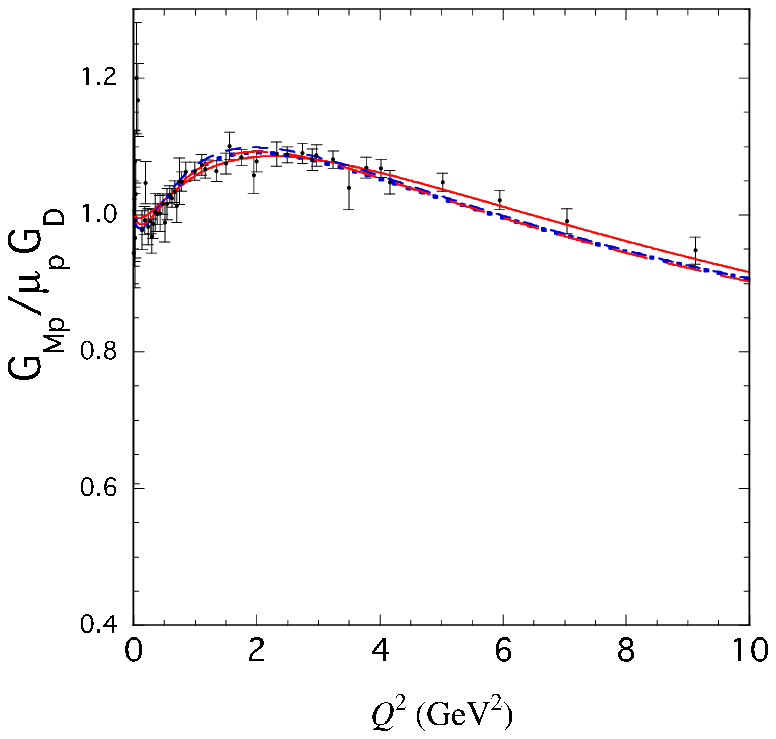}
\includegraphics[width=3.5cm,angle=-90] {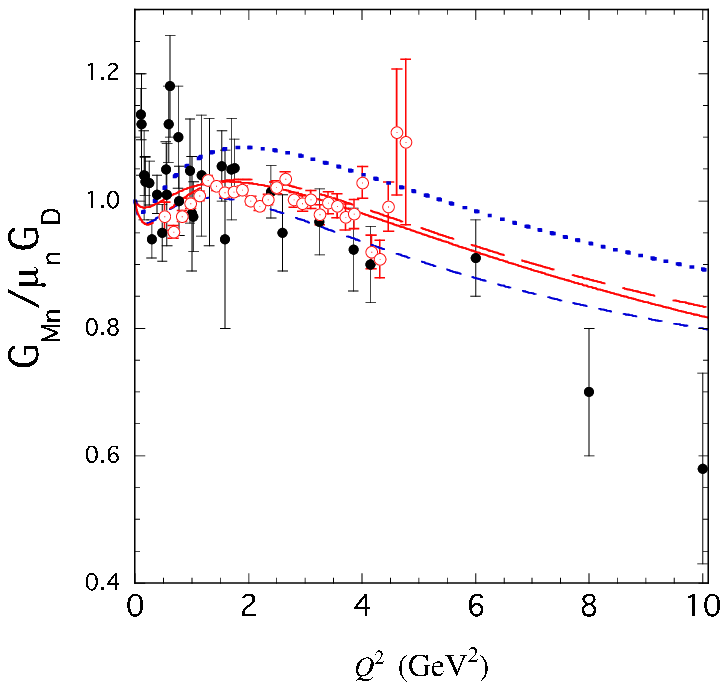}}

\caption{ Nucleon ff's, in the SL region, calculated within the manifestly
covariant spectator model of Ref. \cite{gross}. Dotted line:
 4 parameters,  approximate isospin 
symmetry  in the CQ ff's;
short-dashed line: 5 parameters,   isospin symmetry fully broken; 
long-dashed line: 6 parameters, approximate isospin 
symmetry   + two-pion cut;
 solid line: 9 parameters, isospin symmetry fully broken + two-pion cut.
 (After Ref. \cite{gross})} \label{gross1}\end{figure}

Recently, within the manifestly covariant spectator model, 
  a very  accurate description of the Nucleon ff's
  in the SL region has been achieved \cite{gross}.
   The Nucleon,
 composed of three valence CQ's with  form
factors,  has a center of mass  motion described by a solution of the Dirac equation,
while  the  intrinsic dynamics is given by a  phenomenological S-wave state, consistent with the 
properly symmetrized
covariant spectator formalism. The independence of the intrinsic 
wave
function upon the direction of the relative momentum of an off-mass-shell quark with
respect to an on-mass-shell diquark allows one to construct an internal  pure S-wave Nucleon
state, and therefore to investigate the Nucleon ff's for a Nucleon living in  a well defined orbital angular
momentum state. It is worth noting that the free propagation of 
three quarks is
prevented by a suitable choice of the intrinsic wave function. 
The model has been further  improved by adding a two-pion cut
for a better description of  the ff's at low $Q^2$. In Fig. \ref{gross1}, 
the Nucleon ff's, obtained through  
 fitting procedures that take into account the DIS as well,   are shown.

\begin{figure}[t]
\includegraphics[width=6.2cm,angle=-90] {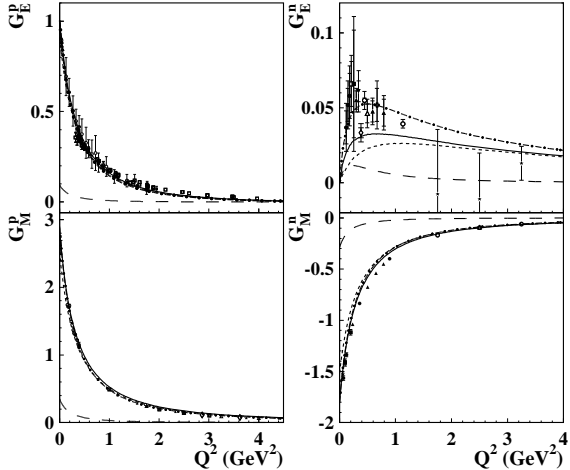}

\caption{Light-front Nucleon ff's in SL region. Long-dashed curve:  
 meson-cloud contribution; dotted curve:  valence-quark
contribution;
solid curve: the sum of the two contributions; dash-dotted
curve:  the total result with a 1\% $S^\prime$-state in
(After Ref. \cite{pasquini}). 
\label{pavia}}
 \end{figure}

Within the Hamiltonian Light-front dynamics, where a meaningful  Fock 
expansion can be considered,
the Pavia group  has devised  a 
CQM + "meson-cloud" approach (i.e., 
$|qqq\rangle +|qqq;q\bar q\rangle$) \cite{pasquini}. In particular, the non
valence term is described by configurations where a   nucleon, or a
Delta,
is coupled to	
$\pi$, $\rho$ or $\omega$, with    extended coupling constants modeled by 
phenomenological vertex functions. 
   The momentum dependence of the nucleon and Delta wave functions is chosen  according
to the  Brodsky-Lepage  prescription  (see, e.g. \cite{brodsky}), while for the mesons a
Gaussian Ansatz is adopted. In  a  
frame where $q^+=q^0+q_z=0$, only $j^+_{em}$ is
necessary, and it is given by a sum of  one-body currents   
for  bare baryons and   mesons. Notably, in the chosen frame,  $j^+_{em}$ is diagonal in
the Fock space. 
After determining i) the two parameters in the pion 
wave function by fitting the
corresponding  ff (for $\rho$ and $\omega$ the same momentum dependence is
assumed) and ii) the cut-off constants in the extended coupling constants,
according to the cloudy bag-model,  one obtains the remaining 3  parameters 
   by fitting the
 theoretical ff's,  to 8 experimental values of  proton and
  neutron  at low $Q^2$ ($\mu_{p(n)}$, $G_A(0)$,
  $G^n_E(0.15~GeV^2)$,$G^p_{E(M)}(0.15~GeV^2)$ and  $G^p_{E(M)}(0.45~GeV^2)$). 
  It is worth
  noting that
 with no $S^\prime$-wave, one obtains $\mu_p=2.87$, $\mu_n=-1.80$, $r_p=0.877~ fm$ 
 and  $r^2_n=-0.064~ fm^2$. In Ref.
  \cite{pasquini}, a
   1\% $S^\prime$ component in
  the Nucleon and Delta wave functions has been also considered, in order to
   allow a better description of 
  $G^n_E$.
  In Fig. \ref{pavia}, the results show that 
  the meson cloud  is significant only for  $Q^2 < 0.5 ~GeV^2$. 
 
\begin{figure}[t]
\includegraphics[width=5.8cm,angle=-90]{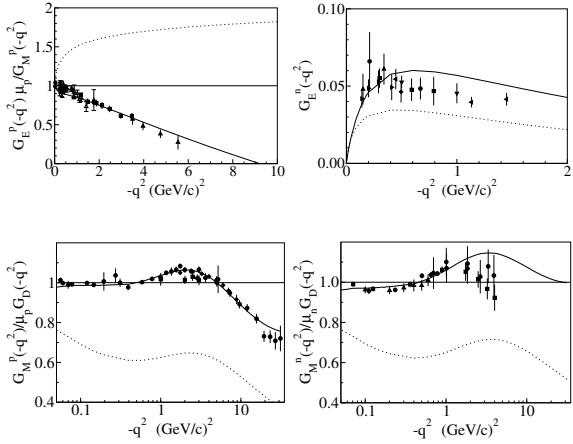}

\caption{SL Nucleon ff's. Solid line: full calculation. 
Dotted line: triangle  contribution only. $G_D=1/(1+|q^2|/0.71)^2$. (After Ref.
\cite{demelo})}
\label{demelof1}\end{figure}

\begin{figure}
\includegraphics[width=2.7cm,angle=-90]{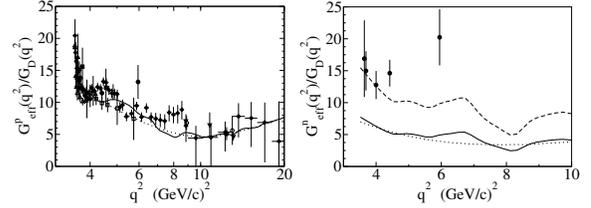}

\caption{TL Nucleon ff's. {Solid line}: full calculation. Dotted line: 
bare contribution only.
Dashed line: full calculation $\times$ 2 (a guide for the eyes).(After Ref.
\cite{demelo})}
\label{demelof2}\end{figure}

 In Ref. \cite{demelo}, a Light-front analysis, already applied
 to   SL
and TL Pion ff \cite{pion}, has been extended to the Nucleon ff's. 
The starting point
is 
the Mandelstam covariant expression for the matrix
 elements of the em  current. It involves: i) the Dirac propagators of the
 quarks, ii) the Nucleon Bethe-Salpeter amplitude (BSA), with a proper 
 Dirac structure and a scalar function describing the four-momenta dependence,
 iii) the quark-photon vertex, to be modeled in order to take into account a
 microscopical Vector Meson Dominance (VMD) contribution as well \cite{pion}. 
 After choosing a frame where $ q^+\ne 0$ (in this frame, $j^+_{em}$ is {\em
 not} diagonal in
the Fock space) , 
 that allows one to
deal with the SL and TL regions on the same footing, one can   easily integrate   
   on $k^-$ (i.e. projecting onto 
 the Light front), if   only the Dirac poles are retained. It turns
 out that
   the different terms that naively one should expect
 from physical considerations, can be straightforwardly singled out (see 
 \cite{demelo}). 
In particular, in the SL
region, besides the standard triangle contribution (or valence term), one has a
 contribution produced by the
pair creation (non valence term or Z-diagram term). This term involves 
both a BSA with 
two on-shell quarks, that can be related to the valence state, and a BSA  
with two off-shell quarks, leading to a
non-valence term in a Fock expansion of the Nucleon state. In the TL region,
there is only the pair contribution.
The quark-photon vertex contains a bare contribution acting 
in the triangle term
  (SL only) and a contribution acting when  the pair production is present,
  namely $Z_b~\times$ a bare term $+~Z_{VMD}~\times$ 
 a VMD term, according to  the decomposition of the photon
  state in bare,  hadronic  (and leptonic) contributions.
The momentum dependence of the Nucleon
BSA in the valence sector has been approximated 
 by a
Nucleon wave function a la Brodsky-Lepage (i.e., pQCD inspired) \cite{brodsky}
with a quark mass  $m_q=200~MeV$ and an overall power 7/2. 
Two parameters are present in the wave function,  one
is 
 fixed through  the anomalous magnetic moments
of proton and neutron, 
($\mu^p_{th}$ = 2.878, Exp. 2.793,  and  $\mu^n_{th}$ =  -1.859, Exp. -1.913), 
and
the other, $p$, that controls the end-point behavior,
 by the overall fitting procedure (see below).
In the non valence sector the momentum dependence  is approximated by
$$
  {\Lambda}^{SL}_{NV}=
  [g_{12}]^{5/2}~g_{N\bar 3}~{(k_1^+ + k_2^+)
 \over  P^{\prime +}_N}  
 ~ \left ({ P^+_N \over k_{\overline {3}}^+ }\right )^{{r}} 
$$
where $g_{AB}=(m_A~m_B) / \left
[\beta^2+M^2_0(A,B)\right]$ and $r$ is analogous to $p$. 
A similar expression holds in the TL region.
The four adjusted parameters, $Z^{IS}_b=Z^{IV}_b =Z^{IV}_{VM}$,   
$Z^{IS}_{VM}$, $p$ and  $r$, are obtained
from the fit of the SL ff's, (even disregarding the proton ratio!) with a
$\chi^2/DOF \sim 1.7$. 
The model, illustrated by Figs. \ref{demelof1} and \ref{demelof2},
 shows that in the SL region the possible zero in $G^p_E$
  is due to the cancellation between the triangle term and 
   the Z-diagram contribution, generated by the
 higher Fock components of the proton state.
 Moreover, in the TL region one can see that for the proton some strength 
 is lacking 
for $q^2 \sim 4.5$ and $\sim 8 ~(GeV/c)^2$ (as for the Pion \cite{pion}), while 
 the neutron remains a challenge.

\section{Conclusions }
 Summarizing,  Lattice data are becoming more and more accurate, 
 but unfortunately only the SL region can be studied, and for the time being in a limited range of
 $q^2$. Phenomenological models, that can address both SL an TL regions, 
 appear still very appealing, and could contaminate each other, e.g. by
 supplying the analytic approaches with ingredients microscopically evaluated.  
In conclusion, one
has a set of   tools, one fundamental and the others phenomenological,
 that can
 positively interact,  allowing 
 to gain a more deep insights in 
the analysis of
the Nucleon ff's, that represent an intriguing challenge for the theory, 
given the highly 
non perturbative nature of the physical quantities involved in the few-GeV
region.

\end{document}